\documentstyle{mn}\onecolumn

\newif\ifAMStwofonts

\def\mincir{\raise -2.truept\hbox{\rlap{\hbox{$\sim$}}\raise5.truept \hbox{$<$}\ }}
\def\mincireq{\hbox{\raise0.5ex\hbox{$<\lower1.06ex\hbox{$\kern-1.07em{\sim}$}$}}}
\def\magcir{\raise-2.truept\hbox{\rlap{\hbox{$\sim$}}\raise5.truept \hbox{$>$}\ }}

\title{Non-thermal emission in the lobes of Fornax\,A}

\author[Persic \& Rephaeli]
       {Massimo Persic$^{1,2,3}$, 
        Yoel Rephaeli$^{4,5}$\\
        $^1$INAF-Trieste Astronomical Observatory, via G.B.\,Tiepolo 11, I-34100 Trieste, Italy \\
        $^2$INFN-Trieste, via A.\,Valerio 2, I-34127 Trieste, Italy \\
        $^3$Physics \& Astronomy Dept., Bologna University, via P.\,Gobetti 93/2, I-40129 Bologna, Italy \\
        $^4$School of Physics \& Astronomy, Tel Aviv University, Tel Aviv 69978, Israel \\
        $^5$Center for Astrophysics and Space Sciences, University of California at 
		San Diego, La Jolla, CA 92093, USA} 
\date{Accepted ... ... ... ... ;
      Received ... ... ... ... ;
      in original form ... ... ... ...}

\pubyear{2019}

\begin{document}

\maketitle

\label{firstpage}

\begin{abstract}

Current measurements of the spectral energy distribution in radio, X-and-$\gamma$-ray provide a sufficiently 
wide basis for determining basic properties of energetic electrons and protons in the extended lobes of the 
radio galaxy Fornax\, A. Of particular interest is establishing observationally, for the first time, the level 
of contribution of energetic protons to the extended emission observed by the {\it Fermi} satellite. Two recent 
studies concluded that the observed $\gamma$-ray emission is unlikely to result from Compton scattering of 
energetic electrons off the optical radiation field in the lobes, and therefore that the emission originates 
from decays of neutral pions produced in interactions of energetic protons with protons in the lobe plasma, 
implying an uncomfortably high proton energy density. However, our exact calculation of the emission by energetic 
electrons in the magnetized lobe plasma leads to the conclusion that all the observed emission can, in fact, be 
accounted for by energetic electrons scattering off the ambient optical radiation field, whose energy density 
(which, based on recent observations, is dominated by emission from the central galaxy NGC\,1316) we calculate 
to be higher than previously estimated. 

\end{abstract}

\begin{keywords}
galaxies: cosmic rays -- galaxies: active -- galaxies: individual: Fornax\,A -- gamma rays: galaxies -- radiation mechanisms: non-thermal
\end{keywords}

\maketitle

\markboth{Persic \& Rephaeli: Non-thermal emission in Fornax\,A lobes}{}

\section{Introduction}

The spectral energy distributions (SED) of energetic particles outside their galactic sources are 
important for determining basic properties of the populations and for assessing the impact of 
the particle interactions in the magnetized plasma of galactic halos and galaxy clusters. 
Knowledge of these distributions is generally quite limited and is based largely on just radio 
observations. Measurements of non-thermal (NT) X-ray and more recently also $\gamma$-ray emission 
from the extended lobes of several nearby radio galaxies provide, for the first time, a tangible 
basis for detailed modeling of the spectral distributions of energetic electrons and protons in 
the lobes. Sampling the SED over these regions by itself, with only limited spatial information, 
yields important insight on the emitting electrons and possibly also on energetic protons whose 
interactions with the ambient plasma may dominate any observed $\sim 100$ MeV emission (from the 
decay neutral pions).

The very luminous nearby radio galaxy Fornax\,A best exemplifies the level of spectral modeling 
currently feasible with supplementary X-and-$\gamma$-ray measurements. This system, located at a 
luminosity distance $D_L = 18.6$\,Mpc (Madore et al. 1999) in the outer region of the Fornax 
cluster, consists of the elliptical galaxy NGC\,1316 and two roughly circular (radius $R \simeq 
11^\prime$) radio lobes centered (at a distance $d \simeq 17.5^\prime$) nearly symmetrically on 
its east and west sides (e.g., Ekers et al. 1983; Isobe et al. 2006; McKinley et al. 2015). The 
galaxy is at the center of a (sub-cluster) group of galaxies, thought to be falling towards the 
cluster center (Drinkwater et al. 2001), which includes the star-forming galaxies NGC\,1310, 
NGC\,1316C, and NGC\,1317.

Detailed modeling of emission from the lobes of Fornax A was carried out by McKinley et al. (2015) 
and Ackermann et al. (2016); these included fitting radio, {\it WMAP}, {\it Planck}, NT emission at 
1 keV, and {\it Fermi} measurements. Whereas the radio and X-ray data points can be fit with a 
truncated electron spectrum, the indication from both analyses was that Compton scattering of the 
(radio-emitting) electrons off the extragalactic background light (EBL) and some assumed level of 
the local optical radiation fields, are too weak to account for the emission observed by {\it Fermi}. 
This led to the conclusion that the 
$\gamma$ emission could only be interpreted as $\pi^0$-decay from {\it pp} interactions; if so, 
this would imply a very high proton energy density in the lobes.

Specifically, the deduced proton energy density was estimated to be $\sim 10^2$ higher than 
the gas thermal energy density, $\sim 1$ eV\,cm$^{-3}$ (assuming $n_H = 3 \cdot 10^{-4}$ 
cm$^{-3}$, for $k_BT = 1$ keV; Seta et al. 2013). To ameliorate this clearly problematic result, 
McKinley et al. (2015) suggested that {\it pp} interactions occur mostly in filaments 
with very high gas over-density and enhanced magnetic field.
This could very well be unrealistic given the implied enhancement also in thermal X-ray emission 
and the enhanced level of radio emission by secondary electrons and positrons produced in charged pion 
decays. Such enhancements would have likely been detected in both radio maps and X-ray images of the 
lobes (e.g., Tashiro et al. 2001, 2009; Isobe et al. 2006).

In an attempt to clarify and possibly remove some of the modeling uncertainty we re-assess 
key aspects of conditions in the Fornax A lobes, and repeat detailed calculations of the 
emission by energetic electrons and protons. We base our analysis on all the available radio, 
X-ray, and $\gamma$-ray measurements, and on newly-published detailed surface-brightness 
distribution of stellar emission from the central galaxy NGC\,1316. The latter emission is 
sufficiently intense to constitute the most dominant optical radiation field in the lobes; 
consequently, the predicted level of $\gamma$-ray emission is significantly higher than in 
previous estimates. In Section 2 we briefly review the observations of NT emission in the 
various bands and the optical radiation field in the lobe region. In Section 3 we describe 
our calculations of the lobe SED model, which is fit to observations in Section 4. We conclude 
with a short discussion in Section 5.

\section{Observations of NT Emission from Fornax\,A}

Fornax\,A lobes have been extensively observed over a wide range of radio and microwave 
frequencies, in the (soft) X-ray band, and at high energies ($\geq 100$ MeV) with the 
{\it Fermi}/LAT. We use the updated measurements of the lobes in all these spectral regions; 
the dataset used in our analysis is specified in Table\,1 where the listed flux densities are the 
combined emission from both the east and west lobes. On sufficiently large scales the radio 
and X-ray emission appears smooth across the lobes, but with east lobe brightness $\sim1/2$ 
that of the west lobe. 

We briefly review the most relevant observations used in our analysis, avoiding many of the 
details that have already been presented in the two recent similar analyses by McKinley et al. 
(2015) and Ackermann et al. (2016). A reasonably accurate description of the optical and IR 
fields in the region of the lobes is important for a correct calculation of the predicted high 
energy X-and-$\gamma$ emission from Compton scattering of energetic electrons (and positrons), 
a new aspect of our analysis that leads to a different conclusion on the possible origin of the 
emission observed by {\it Fermi}.

McKinley et al. (2015) compiled a comprehensive list of previously published radio measurements, 
supplemented by {\it WMAP} and {\it Planck} data in the $22$-$143$\,GHz band, and at 154\,MHz 
(observed with the Murchison Widefield Array). When available, single-lobe flux densities were 
added up to obtain total lobe emission, implying a relatively low contribution from the central 
galaxy. The flux drops sharply at $\nu \magcir 30$\,GHz. 

X-ray observations and spectral analyses were reported by several groups. Early on Feigelson et 
al. (1995), using spatially-unresolved {\it ROSAT} data, argued in favor of a diffuse NT origin 
of the X-ray emission. Kaneda et al. (1995), based on a spectral decomposition of {\it ASCA} 
$1.7$-$7$\,keV data, suggested that $\sim$$1/3$ and $\sim$$1/2$ of the west and east lobe 1\,keV 
flux is NT, with a spectral index consistent with the radio index. This suggestion was confirmed 
by Tashiro et al. (2001) who, based on follow-up {\it ASCA} data, concluded that $\sim$$1/2$ of 
the west lobe flux is NT; Tashiro et al. (2009), analyzing west-lobe {\it Suzaku} data with a 
multi-component spectral model, claimed the $0.5$-$20$\,keV spectrum to be fit by a power-law (PL) 
with an energy index consistent with the radio index. Similarly, Isobe et al. (2006) found that 
east-lobe {\it XMM-Newton} $0.3-6$\,keV data are well fit by a PL with an index consistent with 
the radio index. Following McKinley et al. (2015), we assume that the 1\,keV flux densities 
(Tashiro et al. 2009, Isobe et al. 2006) are indeed NT.

Based on 6.1\,yr of Pass\,8 {\it Fermi}/LAT data, Ackermann et al. (2016) reported extended 
$>$$100$\,MeV lobe emission consistent with the radio lobes' morphology, and point-like emission 
from the radio core. A similar level of emission, based on a smaller dataset and a point-source 
spatial model, had been previously reported by McKinley et al. (2015).


\begin{table}
\caption[] {Emission from the lobes.}

\begin{tabular}{ l  l  l  l  l  l  l}
\hline
\hline

\noalign{\smallskip}
Frequency      &             Energy Flux     & Reference                       & & Frequency     &   Energy Flux            & Reference  \\
log($\nu$/Hz)  &$10^{-12}$erg\,cm$^{-2}$s$^{-1}$ &                             & & log($\nu$/Hz) &$10^{-12}$erg\,cm$^{-2}$s$^{-1}$ &     \\
\noalign{\smallskip}
\hline
\noalign{\smallskip}
7.476          &       $0.634 \pm 0.063$     & Finlay \& Jones 1973         & & 10.515        &  $3.08 \pm 0.185$     & McKinley et al. 2015 \\
7.933          &       $0.814 \pm 0.163$     & Mills et al. 1960            & & 10.607        &  $2.67 \pm 0.240$     & McKinley et al. 2015 \\
8.188          &        $1.16 \pm 0.220$     & McKinley et al. 2015         & & 10.644        &  $2.70 \pm 0.540$     & McKinley et al. 2015 \\
8.611          &        $1.06 \pm 0.106$     & Robertson 1973; Cameron 1971 & & 10.781        &  $2.84 \pm 0.426$     & McKinley et al. 2015 \\
8.778          &        $1.86 \pm 0.465$     & Piddington \& Trent 1956     & & 10.848        &  $2.11 \pm 0.359$     & McKinley et al. 2015 \\
8.926          &        $1.42 \pm 0.128$     & Jones \& McAdam 1972         & & 11.000        &  $1.20 \pm 0.300$     & McKinley et al. 2015 \\ 
9.151          &        $1.77 \pm 0.142$     & Ekers et al. 1983            & & 11.155        &  $0.415 \pm 0.212$    & McKinley et al. 2015 \\
9.179          &        $1.77 \pm 0.199$     & Fomalont et al. 1989         & & 17.38         &  $0.498 \pm 0.100$    & Tashiro et al. 2009; Isobe et al. 2006 \\
9.431          &        $2.65 \pm 0.265$     & Shimmins 1971                & & 22.60         &  $0.7 \pm 0.2$        & Ackermann et al. 2016 \\
9.699          &        $2.74 \pm 0.411$     & Gardner \& Whiteoak 1971     & & 23.15         &  $1.0 \pm 0.2$        & Ackermann et al. 2016 \\
10.353         &        $3.18 \pm 0.159$     & McKinley et al. 2015         & & 23.75         &  $1.1 \pm 0.2$        & Ackermann et al. 2016 \\
10.453         &        $3.10 \pm 0.217$     & McKinley et al. 2015         & & 24.31         &  $0.8 \pm 0.2$        & Ackermann et al. 2016 \\

\noalign{\smallskip}

\hline\end{tabular}
\end{table}
       

\subsection{Optical radiation fields in the lobes}

The superposed radiation field in the lobe region has cosmic and local components.
\medskip

\noindent
{\it a) Cosmic Fields} 

\noindent
{\it i)} The cosmic microwave background (CMB) is described as an undiluted Planckian with $T_{\it CMB} = 2.735$\,K 
and integrated intensity $\nu I_\nu = 960$\,nW\,m$^{-2}$\,sr$^{-1}$ and energy density $u_{\rm CMB}=0.25$\,eV\,cm$^{-3}$
(Cooray 2016); 

\noindent
{\it ii)} The cosmic IR background (CIB), originating from dust-reprocessed starlight integrated over the star formation history 
of galaxies, is described as a diluted Planckian with $T_{\rm CIB} = 29\,^o$K (from its 100\,$\mu$m peak) and integrated intensity 
$\nu I_\nu = 30$\, nW\, m$^{-2}$\,sr$^{-1}$ and energy density $u_{\rm CIB}=7.85 \cdot 10^{-3}$\,eV\,cm$^{-3}$ (Cooray 2016); 

\noindent
{\it iii)} The cosmic optical background (COB), originating from direct starlight integrated over all stars ever formed, is 
described as a diluted Planckian with $T_{\rm COB} = 2900\,^o$K (from its 1\,$\mu$m peak) and integrated intensity $\nu I_\nu = 
24$\,nW\, m$^{-2}$\,sr$^{-1}$, and $u_{\rm COB} = 6.28 \cdot 10^{-3}$\,eV\,cm$^{-3}$ (Cooray 2016). 

The dilution factor, $C_{\rm dil}$, is the ratio of the actual energy density, $u$, to the energy density of an undiluted blackbody 
at the same temperature $T$, namely $u = C_{\rm dil} a T^4$, where $a$ is the Stefan-Boltzmann constant. The dilution factors 
of the cosmic fields are, $C_{\rm CMB} = 1$, $C_{\rm CIB} = 10^{-5.629}$, and $C_{\rm COB} = 10^{-13.726}$.
\medskip

\noindent
{\it b) Local Fields} 

The local radiation fields are dominated by emission from the central galaxy, NGC\,1316, whose SED shows two thermal humps, 
IR and optical. The IR hump peaks at 100\,$\mu$m and has a bolometric luminosity 
\footnote{
The total-IR flux is computed from the {\it IRAS} flux densities at 12, 25, 60 and 100$\mu$m (Golombek et al. 1988) 
using $f_{\rm IR}=1.8 \cdot 10^{-11} (13.48\,f_{12} + 5.16\,f_{25} + 2.58\,f_{60} + f_{100})$\, erg\,cm$^{-2}$s$^{-1}$ 
(Helou et al. 1988). }
$L_{\rm IR} \simeq 1.5 \cdot 10^{43}$\,erg\,s$^{-1}$; its effective temperature is $T_{\rm gal,\,IR} = 29$\,K. The optical hump 
peaks at $1\,\mu$m, corresponding to an effective temperature $T_{\rm gal,\,OPT} = 2900$\,K, and has a bolometric luminosity 
$L_{\rm opt} \sim 1.4 \cdot 10^{45}$\,erg\,s$^{-1}$ from the total $r$-band magnitude (Iodice et al. 2017), converted to bolometric 
magnitude, as detailed below. 

The optical stellar surface-brightness distribution of NGC\,1316, with a half-light radius of $135^{\prime \prime}$, can be 
decomposed into a (nearly) de\,Vaucouleurs spheroid plus an exponential envelope (Iodice et al. 2017). Its deprojection, i.e., 
the emissivity distribution, will be used to model the distribution of the photon energy density. 
\smallskip

\noindent
The de\,Vaucouleurs profile is given by $\sigma(R) = \sigma_0 \, {\rm exp}\left[-\delta\,(R/R_e)^{\alpha}\right]$, where 
$\sigma_0$ is the central surface brightness, $\delta=7.67$, $\alpha = 1/4$, and $R_e$ is the radius encompassing half the 
(integrated) luminosity. The corresponding deprojected profile can be well approximated by the analytical expression 
$\rho(x)=\rho_0\, x^{-\beta} {\rm exp}\left(-x^{\alpha}\right)$ where $x=r\, \delta^4/R_e$, 
$\rho_0 = \sigma_0 \, \delta^4/(2 R_e) \,\Gamma(8)/\Gamma\left((3-\beta) / \alpha)\right)$, and 
$\beta=0.855$ (Mellier \& Mathez 1987). From $r$-band surface photometry measurements (Iodice et al. 2017) $R_e = 87^{\prime 
\prime}$ ($\sim 7.8$\,kpc) and $\mu_{e,\,r} = 20.76$ mag\,arsec$^{-2}$. The latter can be transformed into $b$-band magnitudes 
using $b = r + 1.33\,(g-r) + 0.20$ (Jester et al. 2005), with $g-r = 0.74$ inside the half-light radius (Iodice et al. 2017): 
$\mu_{e,\,b} = 21.94$ mag\,arsec$^{-2}$. The $b$-band bolometric correction, needed to estimate the full-band magnitude, is 
BC$_b = -0.85 - (b-v)$ mag (Buzzoni et al. 2006); with $b-v = 0.84$ (within $2\,R_e$; Cantiello et al. 2013), BC$_b = -1.69$. 
Thus $\mu_{e,\, {\rm bol}} = 20.25$ mag\,arsec$^{-2}$, which corresponds to 
\footnote{
$\mu($mag\,arsec$^{-2}) = M_{\rm bol \sun} + 21.572 -2.5\,$log\,$\sigma(L_{\rm bol \sun}$\,pc$^{-2})$, 
with $M_{\rm bol \sun} = 4.74$ mag and $L_{\rm bol \sun} = 3.83 \cdot 10^{33}$ erg\,s$^{-1}$. 
}
$\sigma_{e,\, {\rm bol}} = 264.9\, L_{{\rm bol}\, \sun}$\,pc$^{-2}$; from $\sigma_0 = e^{7.67} \sigma_e$ we finally derive 
$\sigma_{0,\, {\rm bol}} = 228.3$\, erg\,cm$^{-2}$s$^{-1}$. The resulting volume brightness distribution of the NGC\,1316 
stellar spheroid is 
\begin{eqnarray}
\lefteqn{
\rho_{\rm sph}(r) ~=~ 4.7 \cdot 10^{-21} \, \left({r \over R_e}\right)^{-0.855} \, {\rm exp}\left[-7.67\,\left({r 
\over R_e}\right)^{1/4}\right]  \hspace{0.2cm}  {\rm erg\, cm^{-3}\, s^{-1}} \,.  }
\label{eq:MelMa87}
\end{eqnarray}
\smallskip

\noindent
The deprojection of the exponential profile, $I(R) = I_0 \, {\rm exp}\left[-(R/h)\right]$, where 
$I_0$ is the central surface brightness and $h$ is a characteristic scale, is expressed in terms of a 
modified Bessel function of the second kind, $\rho(r) = I_0 /(\pi \,h) ~ K_0(r/h)$ (Baes \& Gentile 
2011). From Iodice et al. 2017 we get $\mu_{0,\,r} = 22.79$ mag\,arsec$^{-2}$ and $h_r=317^
{\prime \prime}$ ($\sim 28.6$ kpc); however, the stellar envelope is measured to be considerably 
bluer at larger radii (see their Figs.\,4,\,6-left), implying $h_g \sim 360^{\prime \prime}$ ($\sim 
32.5$ kpc). Following the same procedure adopted for the spheroid, $I_{0,\,{\rm bol},} = 0.016$ 
erg\,cm$^{-2}$s$^{-1}$. The resulting emissivity distribution of the NGC\,1316 stellar envelope is 
\begin{eqnarray}
\lefteqn{
\rho_{\rm env}(r) ~=~ 5.2 \cdot 10^{-26} \, \, K_0(r) \hspace{0.2cm}  {\rm erg\, cm^{-3}\, s^{-1}} \,.  }
\label{eq:BaesGent11}
\end{eqnarray}

We can now calculate the galaxy optical photon energy density, $u_{\rm gal\, OPT}$, in the lobes. 
Given that the current X-ray and $\gamma$-ray data are not spatially resolved, we take a volume 
average as a nominal value. The exact calculation accounting for the full light distribution and 
averaging over the volume of the lobes requires a 6D integration, but given the approximate 
nature of our treatment, we compute $u_{gal,\, OPT}$ as a line average along the 
nearest-to-farthest--boundary lobe diameter of the stellar spheroid and envelope components 
\begin{eqnarray}
\lefteqn{
u_{\rm j} = {\rho_0\, \lambda_{\rm j} \over 8 \pi c \rho_s} \,  
\int_0^{2r_s \over \lambda_{\rm j}} 
\int_0^\infty 
\int_0^{2\pi} 
\int_0^\pi 
f_{\rm j} \left( 
\sqrt{
\left[ \rho \sin(\theta) \sin(\phi) + d+ \xi \right]^2 +
\left[ \rho \cos(\theta) \right]^2 + 
\left[ \rho \sin(\theta) \cos(\phi) \right]^2 
} 
\right) \, \sin(\theta) \, d\theta \, d\phi \, d\rho \, d\xi \hspace{0.2cm}  {\rm erg \over cm^{3}}   }
\label{eq:u_lobe}
\end{eqnarray}
where $j$ denotes either component, $\lambda_{\rm j}=R_e, \, h$, and $f_{\rm j}(R/\lambda)$ is given 
by Eqs.\,(\ref{eq:MelMa87}) and (\ref{eq:BaesGent11}). The resulting $u_{\rm j}$ values are $7.500 \cdot 
10^{-14}$ erg\,cm$^{-3}$ (disk) and $4.556 \cdot 10^{-14}$ erg\,cm$^{-3}$ (envelope). 

Finally, $C_{\rm gal\, OPT} = 10^{-12.65}$. A similar calculation yields $C_{\rm gal\, IR} = 10^{-7.00}$. 
Thus, $n(\epsilon)$ includes the CMB, cosmic and local (NGC\,1316) IR/optical radiation fields.

Further contributions to the local optical radiation field should also be assessed. The star-forming 
galaxies NGC\,1310, NGC\,1316C, and NGC\,1317, members of the subcluster that contains Fornax\,A, lie 
close to NGC\,1316. As their estimated projected separations from NGC\,1316 ($\sim 20^\prime$, $30^\prime$, 
$6^\prime$) are comparable to the Fornax\,A size, they are likely located within the lobe region. Their 
cumulative luminosity is relatively modest, $\mincir$15\% of the NGC\,1316 luminosity 
(Iodice et al. 2017).

\section{NT emission in the lobes}

Radio emission in the lobes is by electron synchrotron in a disordered magnetic field whose mean value $B$ 
is taken to be spatially uniform, and X-$\gamma$ emission is by Compton scattering of the electrons by the 
CMB and optical radiations fields. A significant energetic proton component could yield additional radio 
emission by secondary electrons and positrons produced by $\pi^{\pm}$ decays, and $\gamma$-ray emission from 
$\pi^{0}$ decay. The calculations of the emissivities from all these processes are well known and standard, 
and need not be specified here. We will limit our brief treatment here to features in the calculations that 
result from the observational need to explicitly assume (at the outset) that the particle spectral 
distributions are truncated. 

Given the lack of any information about temporal evolution of the lobe emission, it is appropriate 
to represent the spectral electron and proton numbers by time-independent, truncated PL 
distributions in the Lorentz factor $\gamma$ and energy $E$. The spectral (in $\gamma$) number 
of electrons in a lobe is then
\begin{eqnarray}
\lefteqn{
N_e(\gamma) ~=~ N_{e,0} \, \gamma^{-q_e} \hspace{0.25cm}  ... \hspace{0.25cm} 
\gamma_{min} < \gamma <  \gamma_{max}  \,,
}
\label{eq:electron_spectrum}
\end{eqnarray}
in the interval $[\gamma_{min}, \, \gamma_{max}$] with $\gamma_{min}>>1$ and finite $\gamma_{max}$.
Similarly, the spectral (in $E_p$) proton number in a lobe is 
\begin{eqnarray}
\lefteqn{
N_p(E_p) ~=~ N_{p,0} \, E_p^{-q_p} \hspace{0.25cm}  ... \hspace{0.25cm} 
E_p^{min} < E <  E_p^{max} \,,
}
\label{eq:proton_spectrum}
\end{eqnarray}
in the interval $[E_p^{min}, \, E_p^{max}$].

\subsection{Radiative yields of electrons and positrons}

\noindent
Synchrotron emissivity by the above (isotropically distributed) population of electrons is (Blumenthal \& Gould 1970)
\begin{eqnarray}
\lefteqn{
j_s(\nu) ~=~ { \sqrt{3} e^3 B N_{e,0} \over 4 \pi\,m_e c^2} \, 
\int_\Omega \sin \theta \, d\Omega_\theta \, 
\int_{\gamma_{min}}^{\gamma_{max}} \gamma^{-q_e} d\gamma \
\, {\nu \over \nu_c} 
\int_{\nu/\nu_c}^\infty K_{5/3}(\xi) \, d\xi 
\hspace{0.5cm}  {\rm erg}\, {\rm cm}^{-3}\, {\rm s}^{-1}\, {\rm Hz}^{-1} \,, }
\label{eq:synchro_emissivity1}
\end{eqnarray}
where $\nu_c=\nu_0 \gamma^2 \sin \theta$ 
($\nu_0 = {3eB \over 4 \pi m_ec}$ is the cyclotron frequency).
The modified Bessel function of the second kind, $K_{5/3}(\xi)$, has the integral representation
\begin{eqnarray}
\lefteqn{
K_{5/3}(\xi) ~=~ \int_0^\infty e^{-\xi\, \cosh(t)} \cosh\left({5 \over 3} t\right)\, dt \,.}
\label{eq:K53}
\end{eqnarray}
With $x = \nu/\nu_c = \nu/[\nu_0 \gamma^2 \sin \theta ]$ Eq.\,(\ref{eq:synchro_emissivity1}) transforms into 
\begin{eqnarray}
\lefteqn{
j_s(\nu) = N_{e,0} {\sqrt{3} \over 4} {e^3 B \over m_e c^2} \left({\nu \over \nu_0}\right)^{-{q_e-1 \over 2}} 
\int_0^\pi \sin \theta ^{q_e+3 \over 2} 
\int_{\nu \over \nu_0 \gamma_{max}^2 \sin \theta }
^{\nu \over \nu_0 \gamma_{min}^2 \sin \theta }
x^{q_e-1 \over 2}
\int_x^\infty \int_0^\infty e^{-\xi \, \cosh(t)} \cosh\left({5 \over 3} t \right)\, dt \,d\xi\, dx \, d\theta 
\hspace{0.2cm}  {{\rm erg} \over {\rm cm}^3 {\rm s \, Hz}} \,. }
\label{eq:synchro_emissivity_PL1}
\end{eqnarray}
This 4D integral can be readily reduced to a 3D integral 
\begin{eqnarray}
\lefteqn{
j_s(\nu) ~=~ N_{e,0} {\sqrt{3} \over 4} {e^3 B \over m_e c^2} \left({\nu \over \nu_0}\right)^
{-{q_e-1 \over 2}}\int_0^\pi \sin \theta ^{q_e+3 \over 2}
\int_{\nu \over \nu_0 \gamma_{max}^2 \sin \theta }
^{\nu \over \nu_0 \gamma_{min}^2 \sin \theta } x^{q_e-1 \over 2}
\int_0^\infty e^{-x \, \cosh(t)} {\cosh\left({5 \over 3} t \right) \over \cosh\left(t \right)\,} dt \,dx \, 
d\theta\hspace{0.5cm}  {{\rm erg} \over {\rm cm}^3 {\rm s \, Hz}} \,. }
\label{eq:synchro_emissivity_PL2}
\end{eqnarray}

The (differential) number of scattered photons in Compton scattering of electrons from the above distribution by 
a diluted (dilution factor $C_{\rm dil}$) Planckian (temperature $T$) radiation field, 
\begin{eqnarray}
\lefteqn{
n(\epsilon) ~=~ C_{\rm dil}~ {8 \pi \over h^3 c^3}~ {\epsilon^2 \over e^{\epsilon /  k_BT} -1}
\hspace{0.2cm}  {\rm cm^{-3}\, s^{-1}\, erg^{-1}} \,,}
\label{eq:phot_numb_dens1}
\end{eqnarray}
is (Blumenthal \& Gould 1970)
\begin{eqnarray}
\lefteqn{
{ d^2 N_{\gamma, \epsilon} \over dt\, d\epsilon_1 } ~=~
\int_{\epsilon_{min}}^{\epsilon_{max}} 
\int_{ {\rm max} \{ {1 \over 2} \sqrt{\epsilon_1 \over \epsilon},\, \gamma_{min} \} }^
{\gamma_{max}}N_e(\gamma)\, {\pi r_0^2 c \over 2 \gamma^4} \,  {n(\epsilon) \over \epsilon^2} \,
\left( 2\epsilon_1 \, {\rm ln} {\epsilon_1 \over 4\gamma^2 \epsilon} + \epsilon_1 + 4\gamma^2 
\epsilon - {\epsilon_1^2 \over 2\gamma^2 \epsilon} \right)
~ d\gamma \, d\epsilon
\hspace{0.5cm}  {\rm cm^{-3}\, s^{-1}\, GeV^{-1}} \,,}
\label{eq:IC_emissivity2}
\end{eqnarray}
where $\epsilon$ and $\epsilon_1$ are the incident and scattered photon energies, and $r_0 = (e^2/m_ec^2)^2$ is 
the electron classical radius. In Eq.\,(\ref{eq:IC_emissivity2}) we set $\epsilon_{min}=0$ and $\epsilon_{max}/h 
= 10^{12},~ 10^{13}$, and $10^{15}$\,Hz for the CMB, IR, and optical components, respectively. 
\medskip

\subsection{Pion decay yields}

\noindent
Interactions between energetic and (thermal) gas protons lead to the production of $\pi^0$ and $\pi^\pm$, whose 
decays yield $\gamma$\,rays and $e^\pm$, respectively.     
\medskip

The $\pi^0$-decay $\gamma$-ray emissivity is (e.g., Stecker 1971) 
\begin{eqnarray}
\lefteqn{
Q_\gamma(E_\gamma)|_{\pi^0}  ~=~ 
2 \, \int_{E^{min}_{\pi^0}}^{E^{max}_{\pi^0}} \, 
{ Q_{\pi^0}(E_{\pi^0}) \over \sqrt{ E_{\pi^0}^2 - m_{\pi^0}^2 }} \,  dE_{\pi^0} \hspace{0.25cm} 
{\rm cm^{-3}\, s^{-1}\, GeV^{-1}} }
\label{eq:hadr_emissivity1}
\end{eqnarray}
where the factor 2 accounts for the two photons emitted by the $\pi^0$ decay, $E^{min}_{\pi^0} = E_{\gamma} + 
m_{\pi^0}^2 /(4 E_\gamma)$ GeV is the minimum $\pi^0$ energy required to produce a photon of energy $E_\gamma$, 
$E^{max}_{\pi^0}$ is the maximum energy of a neutral pion specified in terms )of the above two masses and the 
maximum proton energy $E^{max}_p$, and 
\begin{eqnarray}
\lefteqn{
Q_{\pi^0}(E_{\pi^0})
 ~=~ 
{4 \over 3} \pi n_H \,\int_{E^{thr}_p(E_{\pi^0})}^{E^{max}_p} \, J_p(E_p) \,
{ d\sigma (E_{\pi^0}, E_p) \over dE_{\pi^0} } \, dE_p 
\hspace{0.25cm}  {\rm cm^{-3}\,s^{-1}\,GeV^{-1}} }
\label{eq:pion_emissivity1}
\end{eqnarray}
is the spectral distribution of the neutral pions. Here, $n_H$ is the (thermal) gas proton density, 
$J_p = {c \over 4 \pi} \, N_p$ is the (isotropic) energetic proton flux, $d\sigma (E_{\pi^0}, E_p) / 
dE_{\pi^0}$ is the differential cross-section for production of neutral pions with energy $E_{\pi^0}$ 
from a collision of a proton with energy $E_p$ with a proton (essentially) at rest, and $E^{thr}_p 
\simeq 1.22$ GeV is the threshold energy for production of a neutral pion with energy $E_{\pi^0}$. 
We use a $\delta$-function approximation (Aharonian \& Atoyan 2000) to the differential cross-section 
by assuming that a fixed average fraction, $\kappa_{\pi^0}$, of the proton kinetic energy, $E_p^{kin}
=E_p - m_p$, is transferred to the neutral pion, i.e., $d\sigma (E_{\pi^0}, E_p) / dE_{\pi^0} \sim 
\delta(E_{\pi^0} - \kappa_{\pi^0} E_p^{kin}) \,\sigma_{pp}(E_p)$ with $\sigma_{pp}$ the total 
cross-section for inelastic {\it p-p} collisions. Then
\begin{eqnarray}
\lefteqn{
Q_{\pi^0}(E_{\pi^0})  
~=~
{c \, n_H \over 3\, \kappa_{\pi^0} } ~ N_p\Big\{m_p + {E_{\pi^0} \over \kappa_{\pi^0}}\Big\} ~ 
\sigma_{pp}\Big\{m_p + {E_{\pi^0} \over \kappa_{\pi^0}}\Big\} \,}
\label{eq:pion_emissivity2}
\end{eqnarray}
(in cm$^{-3}$s$^{-1}$GeV$^{-1}$) where $\kappa_{\pi^0} \simeq 0.17$ in the GeV-TeV region, and 
$\sigma_{pp}(E_p)$ can be analytically approximated (see Eq.\,(79) of Kelner et al. 2006). Noting 
that in the range $5 \mincir E_p/{\rm GeV} \mincir 50$ (relevant to our analysis), $\sigma_{pp}$ 
is roughly constant, $\sigma_{pp,0} \approx 30$\,mbarn, we set $\sigma_{pp} = \sigma_{pp,0}$. The 
resulting $\gamma$-ray photon emissivity (in cm$^{-3}$ s$^{-1}$ GeV$^{-1}$; energies are in GeV) is 
\begin{eqnarray}
\lefteqn{
Q_\gamma(E_\gamma)|_{\pi^0}  ~=~ 
{2 \over 3} \, {c \over \kappa_{\pi^0}} \, n_H \, N_{p,0} \, \sigma_{pp,0}
\int_{E^{min}_{\pi^0}}^{E^{max}_{\pi^0}} { (m_p+E_{\pi^0} / \kappa_{\pi^0})^{-q_p} \over \sqrt{ E_{\pi^0}^2 - m_{\pi^0}^2 }}\, dE_{\pi^0} 
\hspace{0.25cm}  {\rm cm}^{-3}\, {\rm s}^{-1}\, {\rm GeV}^{-1} \,.}
\label{eq:hadr_emissivity2}
\end{eqnarray}
\medskip

The production rate of electrons and positrons is calculated from the $\pi^\pm$ decay rate density (e.g., Stecker 1971),
\begin{eqnarray}
\lefteqn{
Q_{\pi^\pm}(E_{\pi^\pm}) ~=~ 
{2 \over 3} c n_H \int_{E_{\rm thr}} N_p(E_p) ~ f_{{\pi^\pm}, p}(E_p, E_{\pi^\pm}) \, dE_p 
\hspace{0.25cm}  {\rm cm}^{-3}\, {\rm s}^{-1}\, {\rm GeV}^{-1} \,,}
\label{eq:ch_pion_density1}
\end{eqnarray}
where $f_{{\pi^\pm}, p}(E_p, E_{\pi^\pm})$ is the $\pi^\pm$ energy distribution for an incident proton energy $E_p$. 
The $\pi^\pm$ energy distribution can be well approximated by assuming that a constant fraction, $k_{\pi^\pm}$, of the 
proton kinetic energy is transferred to charged pions, so that 
\begin{eqnarray}
\lefteqn{
f_{{\pi^\pm}, p}(E_p, E_{\pi^\pm}) ~=~ \sigma_{pp}(E_p) ~\delta(E_{\pi^\pm} - k_{\pi^\pm} E_p^{\rm kin}) \, . }
\label{eq:ch_pion_distrib}
\end{eqnarray}
For the range of $q_p$ values of interest here, $k_{\pi^\pm}= 0.25$ (Kelner et al. 2006). Substituting 
Eq.\,(\ref{eq:ch_pion_distrib}) in Eq.\,(\ref{eq:ch_pion_density1}) we obtain
\begin{eqnarray}
\lefteqn{
Q_{\pi^\pm}(E_{\pi^\pm}) ~=~ {2 \over 3} \,{c \over k_{\pi^\pm}}\, n_H \, N_{p,0} 
~ \left( m_p+{E_{\pi^\pm} \over k_{\pi^\pm}} \right)^{-q_p} ~
\sigma_{pp}\Big\{m_p+{E_{\pi^\pm} \over k_{\pi^\pm}} \Big\}  	\hspace{0.25cm}  
{\rm cm}^{-3} {\rm s}^{-1} {\rm GeV}^{-1} \,.}
\label{eq:ch_pion_density2}
\end{eqnarray} 
The above equation for the $\pi^\pm$-decay rate density, Eq.\,(\ref{eq:ch_pion_density2}), allows calculation (e.g., 
Ramaty \& Lingenfelter 1966) of the source rate density of the secondary $e^{\pm}$ produced in charged-pion decays,
\begin{eqnarray}
\lefteqn{
Q_{se}(E_e) ~\approx~ {8 \over 3} \,{c \over k_{\pi^\pm}}\, n_H\, N_{p0}\, \sigma_{pp,0} \, 
\left( {4 E_e + m_{\pi^{\pm}} \over \kappa_{\pi^{\pm}} } + m_p \right)^{-q_p}  
\hspace{0.25cm}  {\rm cm}^{-3} {\rm s}^{-1} {\rm GeV}^{-1}  \,.}
\label{eq:secondary_source2}
\end{eqnarray}

The corresponding steady-state distribution, $N_{se}(\gamma)$, can be calculated via $d \left[b(\gamma) N_{se}(\gamma) \right] 
/ d\gamma = Q_{se}(\gamma)$ -- where, assuming transport losses to be negligible, $b(\gamma)$ is the radiative energy loss term 
(e.g., Rephaeli 1979). The latter is written as $b(\gamma) = b_0 + b_1(\gamma) + b_2(\gamma)$ (in s$^{-1}$), where $b_0(\gamma) 
= 1.2 \cdot 10^{-12} n_e \left[ 1 + {\rm ln} \, (\gamma/n_e) / 84 \right]$ is the loss rate by electronic excitations in ionized 
gas, $b_1(\gamma) = 10^{-15}\gamma \,n_e$ is the loss rate by electron bremsstrahlung in ionized gas, and $b_2(\gamma) = 1.3 \cdot 
10^{-9} \gamma^2 \, (B^2 + 8\pi \rho_r)$ is the Compton-synchrotron loss rate. In these expressions $n_e$ is the thermal electron 
density and $\rho_r$ is the ambient radiation field energy density. The secondary electron density is then 
\begin{eqnarray}
\lefteqn{
N_{se}(\gamma) ~=~ { \int Q_{se}(\gamma) \, d\gamma  \over  b(\gamma) }   \hspace{0.25cm} 
{\rm cm}^{-3} \,,}
\label{eq:ss_se_spectrum1}
\end{eqnarray}
where 
\begin{eqnarray}
\lefteqn{
\int Q_{se}(\gamma) \, d\gamma 
~\approx~ 
{2 \over 3} \,{c \over q_p -1}\, n_H\, N_{p0} \, \sigma_{pp,0} \, m_e^{1-q_p}
\left({4 \over \kappa_{\pi^{\pm}} } \gamma + { m_{\pi^{\pm}} + \kappa_{\pi^{\pm}} m_p \over \kappa_{\pi^{\pm}} m_e} \right)^{-q_p+1} 
 	\hspace{0.25cm}  {\rm cm}^{-3} {\rm s}^{-1} \,. }
\label{eq:integrated_secondary_source}
\end{eqnarray}
The resulting $N_{se}(\gamma)$ can be approximated by 
\begin{eqnarray}
\lefteqn{
N_{se}^{\rm fit}(\gamma) ~=~ N_{se,0} \left[1+\left({\gamma \over \gamma_f}\right)^\tau 
\right]^{-(q_p+1)}   \,. } 
\label{eq:fit_ss_se_spectrum2}
\end{eqnarray}
where $\tau$ is a (fit) parameter specified below (see section 5).

%
\begin{figure}
\vspace{9.0cm}
\includegraphics{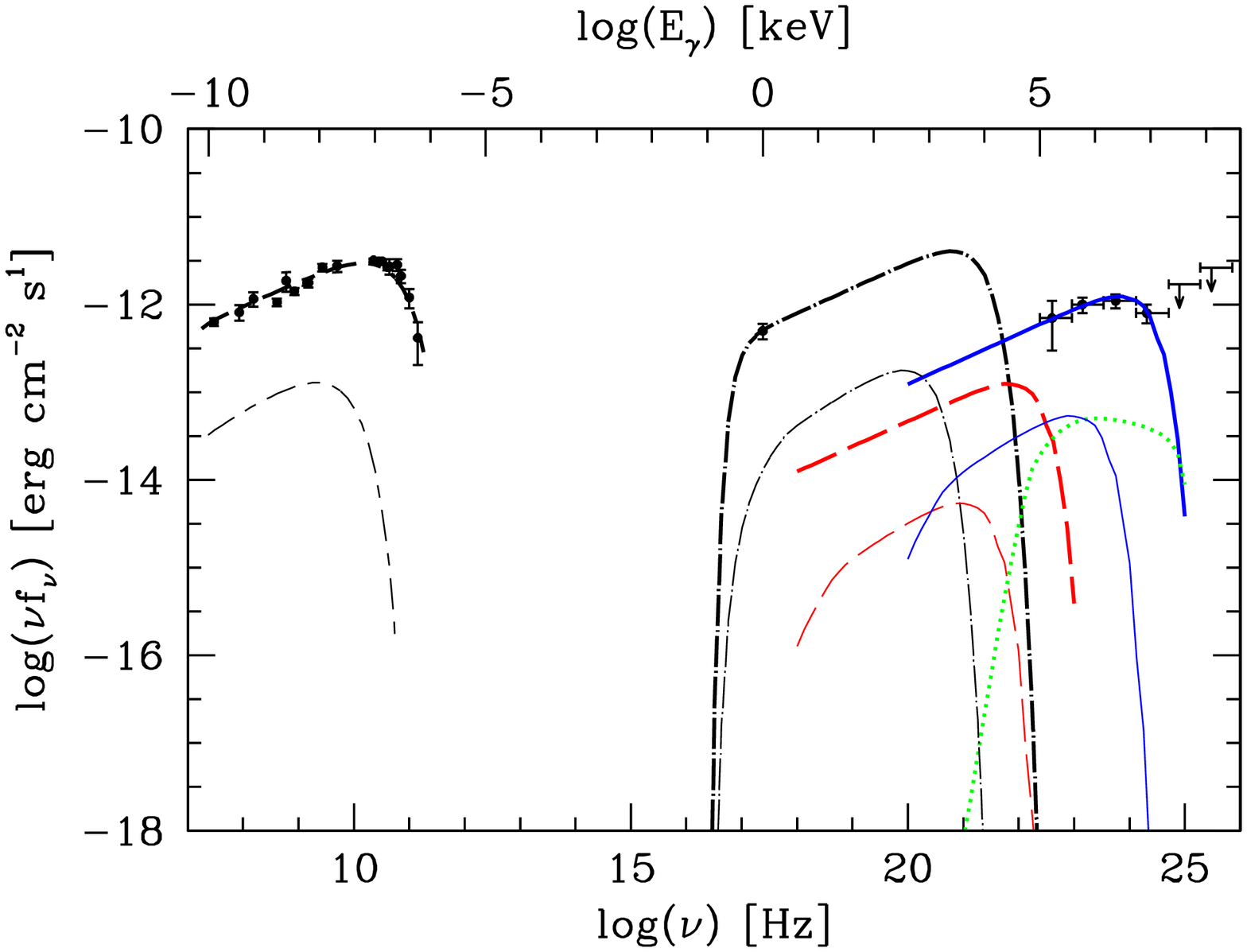}
\includegraphics{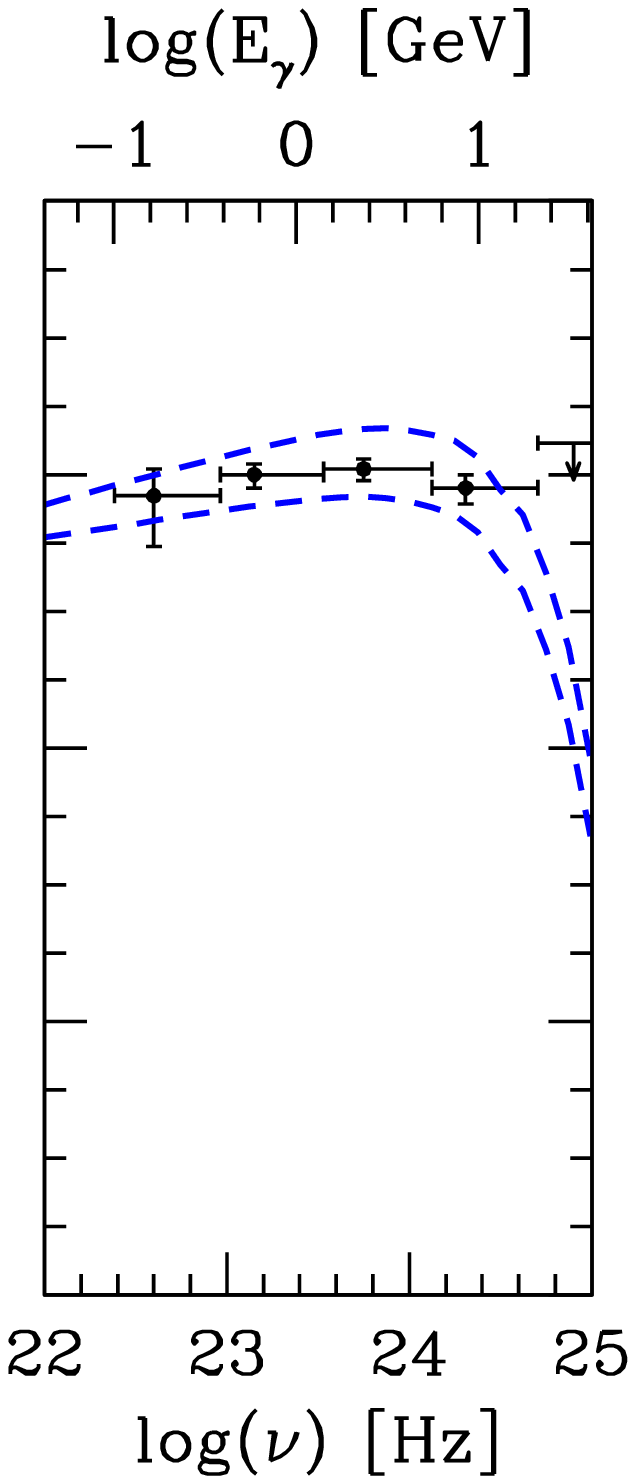}
\caption{ 
{\it Left:} broad-band SED of the Fornax\,A lobes. Data points (Table\,1) are shown by dots.
Emission components are:
synchrotron, black short/long-dashed curves; 
Compton/CMB, black dotted--long-dashed curves;
Compton/IR, red long-dashed curves; 
Compton/OPT, blue solid curves; 
pionic, green dotted curve.
Thinner curves depict secondary-electron emissions.
{\it Right:} dashed curves bracket the $95$\% confidence level of the predicted Compton/OPT emission.
}
\label{fig:SED}
\end{figure}

\section{Modelling the lobe SED} 

As has already been deduced in previous analyses (McKinley et al. 2015; Ackermann et al. 2016), 
the energy range of electrons emitting the observed radio and-X-ray radiation is relatively narrow, 
as is clear from the steep drop of the flux at high radio frequencies, and from the level and 
spectral shape of the emission around $\sim 1$\,keV. 
The exact emission level near this energy is pivotal to our analysis. X-ray observations were 
reported by several groups, starting with spatially-unresolved {\it ROSAT} measurements by 
Feigelson et al. (1995). From spectral analysis of {\it ASCA} $1.7 - 7$\,keV data Kaneda et al. 
(1995) deduced that the emission had a NT component (with a PL index similar to the radio 
index); specifically, they estimated that $\sim 1/3$ and $\sim 1/2$ of 1\,keV flux from the west 
and east lobes, respectively, was NT; this was later confirmed by Tashiro et al. (2001). Similarly, 
Isobe et al. (2006) found that east-lobe {\it XMM-Newton} $0.3-6$\,keV data are well fit by a 
PL with an index consistent with the radio index. Additionally, from analysis of {\it Suzaku} 
data Tashiro et al. (2009) estimated that $\sim 1/2$ of the west lobe flux to be NT. 
Assuming all the measured flux at 1\,keV originates in Compton scattering off the CMB yields the 
overall normalization factor, $N_{e,0} \simeq 8 \cdot 10^{64}$, and the lower energy limit, 
$\gamma_{min}\simeq 200$. 

Fitting the predicted synchrotron flux to measurements in the $0.03 - 143$ GHz yields the spectral 
index of the electron distribution, $q_e = 2.43$, and the upper energy cutoff, $\gamma_{\max} = 7 
\cdot 10^4$. The mean magnetic field strength in the lobes is computed using (e.g., Eq.\,4.53 in 
Tucker 1975) 
\begin{eqnarray}
\lefteqn{
B ~=~ \left[ 2.47 \cdot 10^{-19} ~(5.25 \cdot 10^3)^{(q_e-1)/2} ~ T_{\rm CMB}^{(q_e+5)/2} 
~{b(q_e) \over a(q_e)} ~\left[ {F_c \over F_s} \left( {\nu_c \over \nu_s} \right)^{(q_e-1)/2} 
\right]^{-1} \right]^{2/(q_e+1)} {\rm gauss} \,. }
\label{eq:B_synchro_IC}
\end{eqnarray}
Eq.\,(\ref{eq:B_synchro_IC}) applies when the synchrotron and Compton fluxes are emitted from the 
same region by electrons with untruncated PL spectrum. Substituting the measured 5\,GHz and 1\,keV 
flux densities for $F_s$ and $F_c$, respectively, Eq.\,(\ref{eq:B_synchro_IC}) returns $B = 2.64\,\mu$G. 
The value we find is only slightly different, $B = 2.9\, \mu$G; the $\sim$$10$\% discrepancy mostly 
stems from the scatter of radio data points about the synchrotron curve.

With the electron distribution fully determined, the Compton yields from scattering off the above 
specified IR and OPT radiation fields can now be calculated; the results for the full SED are shown 
in Fig.\,(\ref{fig:SED}) together with the radio, X-ray, and $\gamma$-ray measurements. The main 
current interest in the predicted Compton yields is the emission in the range $0.1 - 100$ GeV probed 
by {\it Fermi}-LAT; the fact that our predicted emission in this range matches well the measurements 
is an important outcome of our analysis that is essentially a direct consequence of our full accounting 
for the optical radiation field in the lobes which yielded a higher energy density than estimated by 
Ackermann et al. (2016). As specified in the previous section, in our detailed estimate of the optical 
radiation field in the lobes we determined that this field is dominated by the stellar emission of the 
central galaxy, NGC\,1316. This conclusion is based on recent mapping of the extended surface brightness 
distribution of this galaxy by Iodice et al. (2017). 

The energetic proton density in the lobes is likely to be quite significant and the energy density 
higher than that of the electrons; however, this does not necessarily imply that the associated 
$\pi^{0}$-decay yield can exceed the above estimated level from Compton scattering of primary 
electrons off the local optical radiation field. Assuming a proton energy distribution, with 
$q_p=2.2$ and a theoretically-motivated nominal proton-to-electron energy-density ratio $\zeta 
=\rho_{p}/\rho_{e} = 50$ (see below), we can calculate the secondary electron and $\pi^{0}$-decay 
yields in all the above spectral bands. These emissions are all well below the corresponding levels 
by primary electrons. Thus, we can only place a $\sim 30\%$ nominal upper limit on the pionic 
contribution to the flux in the {\it Fermi}-LAT band, a limit that roughly reflects the modeling 
uncertainties in our analysis and the level of precision of the current {\it Fermi}-LAT data.

\section{Discussion} 

Detailed modeling of emission in radio lobes is clearly needed in order to determine key properties 
of energetic particles, magnetic and radiation fields, and for assessing the impact of AGN jets on 
their intergalactic or intracluster environment. In particular, fitting model predictions to 
measurements of the lobe SED is critical for a robust determination of the relative significance of 
energetic protons when gauged by their radiative yields from interactions with ambient gas. By virtue 
of proximity, brightness of its lobes, and adequate multi-spectral observations, Fornax\,A is one of 
the most suitable for such a study. 

In our analysis of the broad-band SED of the Fornax\,A lobes we use the simplest truncated-PL spectral 
distribution, the most constraining SED dataset currently available, and a recently published, sufficiently 
precise NGC\,1316 light distribution. Assuming the X-ray flux originates from electron Compton scattering 
off the CMB, our main result is that the related flux from Compton scattering off the optical radiation 
field, which is dominated by the central galaxy, is consistent with the {\it Fermi}-LAT measurements. Thus, 
there is no apparent need for an additional pionic component (at a level comparable to current observations). 
This conclusion would only be strengthened if account is taken of the estimated $\mincir 15$\% contribution 
to the observed emission by NGC\,1316 (Ackermann et al. 2016), and an additional $\mincir 15$\% enhancement 
of the intra-lobe optical light by emission from nearby star-forming galaxies. In light of this, only an upper 
limit can be set on a (likely) pionic component, in contrast with the conclusion reached by McKinley et al. 
(2015) and Ackermann et al. (2016). 

Limits on the ambient proton spectral parameters can be set by selecting the proton distribution's spectral 
index and maximum energy to be close to the values that would be implied from fitting a pionic component to 
the {\it Fermi}-LAT data; this yields $q_p=2.2$ and $E_p^{max}= 50$ GeV, respectively. The value of the 
index is in the range of what is theoretically predicted. Assuming a nominal p/e energy density ratio of $50$, 
as appropriate for an electrically neutral NT plasma for the deduced value of $q_e$ and assumed $q_p$ (Persic 
\& Rephaeli 2014), then $N_{p0} = 1.7 \cdot 10^{62}$. This would imply a total NT proton energy of $7.2 \cdot 
10^{58}$ erg, roughly two orders of magnitude lower than what would be required if the measured $\gamma$-ray 
emission were of pionic origin (McKinley et al. 2015; Ackermann et al. 2016).

This low-level normalization implies also a correspondingly low secondary electron (and positron) contribution 
to the observed emission; thus, the exact values of the proton and secondary-$e^\pm$ spectral indices are 
of little practical interest. More relevant is the fact that the deduced spectral index, $q_e \sim 2.4$, is 
significantly lower than the expected value for a population that had aged as result of efficient Compton-synchrotron 
energy losses -- whose characteristic time, $t_{\rm CS} = \gamma / b_2(\gamma) \sim 0.08$\,Gyr (estimated using 
the relevant photon and magnetic field energy densities and $\gamma = 1.35 \cdot 10^4$, i.e., the likely Lorentz 
factor of an electron emitting at $\nu_c = 1.4$\,GHz in the deduced lobe magnetic field, derived from the expression 
for $\nu_c$ with $<$$\sin \theta$$> = 2/\pi$), is shorter than the estimated age of the lobes, $t_{\rm lobe} 
\sim 0.4$\,Gyr (Lanz et al. 2010). This is perhaps due to lack of information on the particle injection (by the 
jet) and propagation mode and related spectro-spatial aspects, or perhaps a consequence of efficient re-acceleration 
that can flatten pre-existing NT particle spectra (e.g., Bell 1978; Wandel et al. 1987; Seo \& Ptuskin 1994) within the lobes. 
Given these uncertainties, taking $\tau = 3/4$ in Eq.\,(\ref{eq:fit_ss_se_spectrum2}) effectively implies a steady-state 
secondary $e^\pm$ index of (asymptotically) 2.4, same as that of the primary electron spectrum. The secondary electron 
maximum energy, $\sim {1 \over 4}E_p^{max}$ 
corresponds to $\gamma_{max} \sim 2.5 \cdot 10^4$.

To assess the accuracy of our quantitative results we focus on the impact of the main observational and modeling 
uncertainties. Key parameters are the electron number normalization and endpoints of the energy range which were 
determined from the radio and X-ray data. Specifically, the values of $N_{e,0}$ and $\gamma_{min}$ were deduced 
from the measured 1\,keV flux, interpreted to be a consequence of Compton scattering off the CMB. While there is 
appreciable uncertainty in the level of the NT emission at this energy, which was determined by Tashiro et al. 
(2009) and Isobe et al. (2006), it largely impacts the modeled level of the radio spectrum, whose fit to the radio 
data yields the value of the mean magnetic field. Therefore, the resulting uncertainty is essentially reflected in 
the latter quantity whose exact value (in the range $B \sim 1-3\, \mu$G in all previous analyses) is clearly not 
that important (also because it constitutes a nominal volume-averaged value across the lobes). The substantial 
uncertainty in the spectral shape of the X-ray data results in some level of parameter degeneracy; e.g., the 
combination $N_{e,0} = 1.1 \cdot 10^{65}$, $\gamma_{\min}=700$ is consistent with the 1\,keV flux density, but this 
higher $\gamma_{min}$ causes a steeper rise of the predicted flux than suggested by the data. Also, the value of 
$\gamma_{max}$, constrained by the high-frequency turnover of the radio spectrum, is affected by the observational 
error in the value of $q_e$, whose relative level (at $1\,\sigma$ confidence level) is estimated to be $\sim 4$\%.

A source of (mostly) modeling uncertainty results from monochromatic to bolometric flux conversion required in order 
to compute the dilution factor of the optical radiation field sourced by NGC\,1316. This involves converting $r$ to 
$b$ and (then) to bolometric magnitudes based on the adopted stellar population synthesis model. The CIB and COB 
intensities are known with a $\sim 30$\% uncertainty (see Cooray 2016). 

Important are also the relatively large error intervals of the {\it Fermi}-LAT data (below $\sim 20$ GeV). With the 
best-fit value of the spectral index determined from the radio data, $\alpha \simeq 0.71\pm 0.04$ ($1\,\sigma$ 
uncertainty), the propagated uncertainty in the predicted Compton flux in the {\it Fermi}-LAT range is $\sim 15$\%. 

A similar joint analysis of the multi-spectral emission from the giant radio lobes of the nearby radiogalaxy Centaurus\,A, 
the first extended extragalactic regions detected by {\it Fermi}-LAT (Abdo et al. 2010; Yang et al. 2012), is clearly of 
much interest. The lack of unambiguous evidence for NT X-ray emission from the lobes, perhaps due to their large angular 
size and complex X-ray morphology (e.g., Schreier et al. 1979; Hardcastle et al. 2009 and references therein), does not 
allow a definite conclusion on the origin of the measured $\gamma$-ray emission. Assuming that the observed low energy 
($\magcir 60$ MeV) emission originates mostly from Compton scattering off the CMB and EBL, allows calibration of the 
electron spectrum in a purely leptonic model (Abdo et al. 2010
\footnote{The optical emission from the host galaxy, NGC\,5128, 
was estimated to negligibly contribute to the Compton yield.}
). However, the improved spatial resolution attained in more recent radio measurements (Sun et al. 2016) indicates a 
possible magnetic enhancement at the edge of the south lobe, and thus leads to a lower electron number normalization 
that lowers the likelihood of the leptonic origin of the $\gamma$-ray emission. As previously suggested for 
Fornax\,A, a combined lepto-hadronic model seems to require an unrealistically high proton energy density (Sun 
et al. 2016).
 
Unequivocal observational evidence for energetic protons in the lobes of radio galaxies is of great interest also for 
an improved understanding of the origin of extended radio halos in clusters. In a first detailed assessment of galactic 
energetic proton and electrons sources in clusters (Rephaeli \& Sadeh 2016), it was assumed that electrons diffuse out 
of radio and star-forming galaxies, and because of the lack of clear evidence for an appreciable proton component in the 
lobes of radio galaxies, their (secondary electron) contribution to the radio halo emission was conservatively ignored. 
The extended distribution of star-forming galaxies, whose relative fraction increases with distance from the cluster 
center, is important in order to account for the large size of radio halos. This approach was applied to conditions in 
the Coma cluster, where the number of star-forming galaxies was estimated from the total blue luminosity of the cluster, 
and only the two central powerful radio galaxies were included as electron sources. It was found that for reasonable 
models of the gas density and magnetic field spatial profiles, the predicted profile of the combined radio emission from 
primary and secondary electrons is roughly consistent with that deduced from current measurements of the Coma halo. 
However, the level of radio emission was predicted to be appreciably lower than the measured emission, suggesting that 
there could be additional particle sources, such as AGN and (typically many) lower luminosity radio galaxies. Clearly, 
therefore, quantitative evidence for energetic protons in radio lobes would have important implications also on our 
understanding of the origin of cluster radio halos.
\medskip

\noindent
{\it Acknowledgement.} 
We used the NASA/IPAC Extagalactic Database (NED) which is operated by the Jet Propulsion Laboratory, California Institute 
of Technology, under contract with the National Aeronautics and Space Administration. This paper is dedicated to the memory 
of our Bologna University colleague, Prof. Giorgio Palumbo.

\end{document}